\documentstyle[11pt,newpasp,twoside,epsf]{article} 

\newcommand{\vhbb}{$\Delta F555W_{\rm HB}^{\rm Bump}\,$} 
\def\edcomment#1{\iffalse\marginpar{\raggedright\sl#1\/}\else\relax\fi} 
\reversemarginpar 

\begin{document} 
\title{The Red Giant Branch Bump}

\author{Marco Riello, Giampaolo Piotto, Alejandra Recio-Blanco} 
\affil{Dipartimento di Astronomia, Universit\`a di Padova, 
Vicolo dell'Osservatorio, 2, I-35122, Padova, Italy}
\author{Santi Cassisi} 
\affil{INAF-Osservatorio Astronomico di Collurania, Via M. Maggini,
I-64100, Teramo, Italy} 
\author{Maurizio Salaris} 
\affil{Astrophysics Research Institute, Liverpool John Moores 
University, Twelve Quays House, Egerton Wharf, Birkenhead L41 1LD, UK} 

\begin{abstract} 
We present a comparison between theoretical models and the observed
magnitude difference between the horizontal branch and the red giant
branch bump for a sample of 53 clusters. We find a general agreement,
though some discrepancy is still present at the two extremes of the
metallicity range of globular clusters.
\end{abstract}

\section{The red giant branch bump}
The \lq{bump}\rq\ is an intrinsic feature of the red giant branch
(RGB) luminosity function (LF) of globular clusters (GC). It appears
as a peak in the differential LF, and as a change in the slope of the
cumulative LF. The bump originates when the H-burning shell crosses
the chemical discontinuity left over by the convective envelope soon
after the first dredge-up.  So, a comparison between the theoretical
and observed RGB bump location is a powerful tool for checking the
capability of stellar models to finely predict the maximum inward
extension of outer convection during the first dredge-up.  In order to
overcome the uncertainties related the GC distance scale, the location
of the RGB bump is linked to the horizontal branch (HB), and the
parameter $\Delta V_{\rm HB}^{\rm Bump}\,$, (i.e., the difference in
visual magnitude between the RGB bump and the HB stars
within the RR Lyrae instability strip) is commonly used.

Recently, an exhaustive comparison between theory and a large sample
of GCs has been performed by Zoccali et al. (1999 Z99) and Ferraro et
al. (1999). Here, we have doubled the Z99 sample (to 53 clusters),
using the same HST snapshot database (Piotto et al. 2002). At variance
with Z99, and in order to minimize the problems coming from the
trasformation of the models from the theoretical to the observational
plane, here we have worked directly in the HST F555W and F439W flight
system. The analysis procedures are as in Z99. Figure 1 shows the
comparison between the observed \vhbb and the models (Cassisi \&
Salaris 1997) for three different cluster ages.  The trend of
the empirical data is well reproduced by standard stellar models.
In particular,  the models
predict a significant change in the slope of the
\vhbb - [M/H] relation around $[M/H]\approx-0.5$ dex that is
confirmed by the observations.
Some discrepancy is still present at the two extremes of the
metallicity range of GCs.

\begin{figure}
\vspace{-.1in}
\plotfiddle{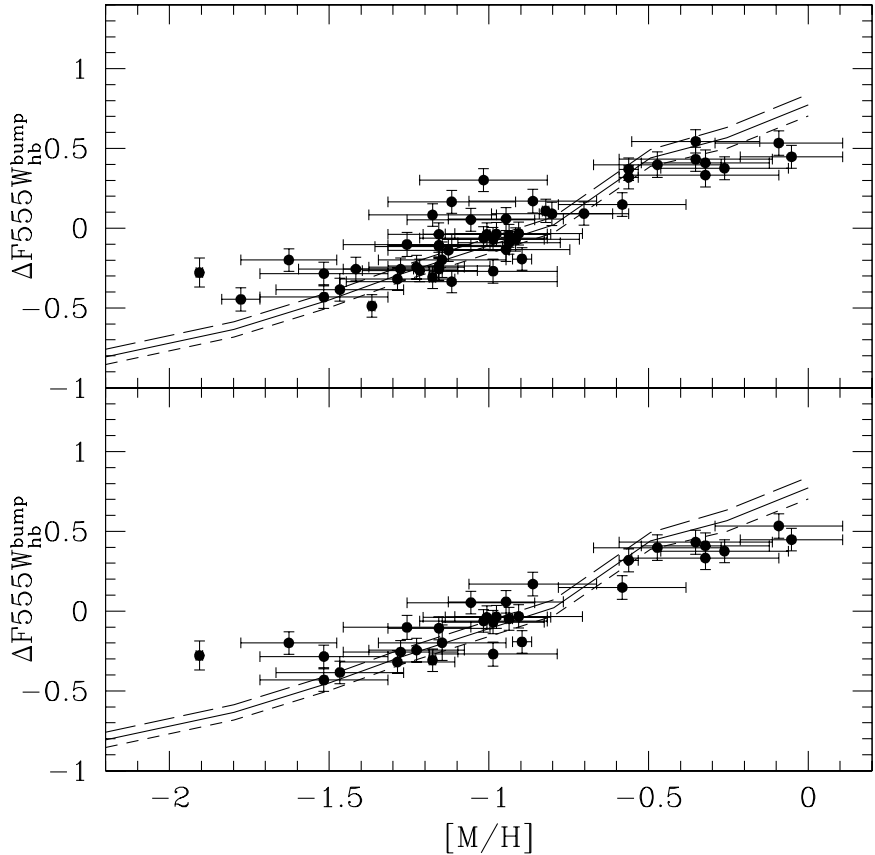}{10.8cm}{0}{60}{60}{-180}{-100}
\caption{Comparison between the observed \vhbb and the models for three ages 
(12, 14, 16 Gyr). We used the Carretta and Gratton (1997) metallicity
scale, modified as in Z99, in order to take into account the
$\alpha$-enhancement. The upper panel shows the entire sample (53
GCs). The lower panel shows clusters which have at least 800 stars in
the RGB (32 GCs).}
\end{figure}

\end{document}